\documentclass[a4paper]{jpconf}
\usepackage{graphicx}
\usepackage{cite}

\begin{document}
\title{Effect of Acoustics on Droplet Grouping Behaviour in a Single Stream of Droplets}

\author[label1]{M. Kumar$^{*1}$, V. Vaikuntanathan$^2$, M. Ibach$^1$, A. Arad$^3$, R. Bar-On$^4$, B. Weigand$^1$, D. Katoshevski$^3$ and J. B. Greenberg$^4$}

\address{$^1$Institute of Aerospace Thermodynamics (ITLR), University of Stuttgart, Stuttgart, Germany}
\address{$^2$Department of Mechanical Engineering, Shiv Nadar Institution of Eminence (deemed to be University), UP, India}
\address{$^3$Department of Civil and Environmental Engineering, Ben-Gurion University of the Negev, Beer-Sheva, Israel}
\address{$^4$Aerospace Engineering, Technion – Israel Institute of Technology, Haifa, Israel}

\ead{manish.kumar@itlr.uni-stuttgart.de}

\begin{abstract}
Droplet and particle grouping can be influenced by applying an acoustic field and have practical applications such as particle scavenging and aerosol filters of engine exhaust and air purifiers. The present work experimentally investigates the influence of a standing acoustic wave on a single stream of droplets. The experimental setup consists of an acoustic transducer and a reflector plate through which the droplet stream passes in the presence or absence of an external pressure field generated by a standing acoustic wave. A droplet stream is generated with the help of a nozzle connected to a pressurized working fluid supply and piezoelectric transducer to control the spacing between droplets. The effect of the acoustic pressure field on the droplet stream generated by the nozzle operated at different piezoelectric excitation frequencies and fluid pressures is investigated. Droplet stream characteristics at every nozzle excitation frequency are observed with a high-speed camera when the acoustic field is switched OFF and ON. The competing effect of nozzle excitation frequency and acoustic field is observed. At lower nozzle frequencies, the nozzle generates an unstable stream of droplets having different sizes and spacings between them. When the acoustic field is applied at these lower frequencies, the stream of droplets becomes organized, and in some cases, it becomes equispaced and of the same size. However, an opposite behavior is observed at higher frequencies. In these cases, as the acoustic field is applied, an equispaced mono-disperse droplet stream becomes unstable due to the coalescence of droplets within the stream.
\end{abstract}

\section{Introduction}

Droplet grouping refers to the tendency of droplets that were initially separated from each other to move closer and coalesce. This phenomenon is observed in various applications, spanning from precise coating technologies to the vaporization and combustion of fuel droplets in fuel sprays. Furthermore, research on droplet dynamics and grouping has gained increased significance, especially in recent times, as droplets can carry viruses associated with life-threatening diseases like COVID-19. Numerous studies in the literature have explored the phenomenon of 'clustering' of droplets within an ambient medium, primarily air, across diverse scenarios ranging from laboratory-based liquid sprays \cite{heinlein2006,sahu2018} to atmospheric clouds \cite{kostinski2001}. These investigations have predominantly focused on examining the characteristics of 'droplet clusters,' including the spatial and temporal distribution of droplet number density, and the impact of such clusters on evaporation and condensation processes. However, the precise process leading to the formation of these clusters and the underlying physical mechanisms remain not fully comprehended. In light of this, studies on 'droplet grouping' gain relevance, as they delve into the tendency and associated mechanisms governing the formation of droplet groups or clusters.

In experiments conducted at high Reynolds numbers \cite{saw2008}, a distinct phenomenon of inertial clustering was observed in regions characterized by maximum fluid acceleration and vorticity. Notably, the degree of clustering exhibited a monotonic increase with the droplet Stokes number. In an earlier study \cite{collins2004}, through Direct Numerical Simulations (DNS), it was demonstrated that in a turbulent flow, denser particles or droplets with finite inertia exhibit a tendency to cluster within regions characterized by low vorticity in the lighter surrounding medium. In another study \cite{heinlein2006}, utilizing a multi-point statistical model applied to Phase Doppler Anemometer (PDA) measurements, droplet clustering emerged as a distinct illustration of unsteadiness in sprays from pressure and air-assisted atomizers. Nevertheless, this study did not identify any significant correlation between droplet size and the occurrence of droplet clusters. Additionally, the regions where droplet clusters were observed were linked to low velocities, attributed to a larger drag force experienced by these clusters of droplets. In three-dimensional particle tracking experiments \cite{yavuz2018} involving droplets in turbulent airflow, inertial clustering was observed, accompanied by an additional occurrence of small-scale extreme clustering. The latter phenomenon was elucidated using a Stokes flow description of two spheres, incorporating hydrodynamic interaction and a perturbative term to account for finite inertial effects. In a recent study, Vaikuntanathan et al. \cite{visakh2022} systematically investigated the droplet grouping phenomenon and its mechanism in a stream of droplets. They observed that the leading droplet experiences a higher drag force compared to the trailing one, which forces droplets to group together and coalesce eventually. A DNS study conducted by Ibach et al. \cite{ibach2022} on droplet grouping in a stream of droplets corroborated the same. deBotton et al. \cite{de2022} analytically investigated the grouping of two droplets of dissimilar sizes and reported that the larger droplet approaches slowly to the other droplet compared to a group of same-size droplets.

In the literature, it has been reported that droplet and particle grouping can be influenced by applying an acoustic field and have practical applications such as particle scavenging and aerosol filters of engine exhaust and air purifiers \cite{katoshevski2010,visakh2021a,ma2017,ran2014}. Previously, several studies \cite{achury2017,sujith2005,katoshevski2006,ficuciello2017} have been done to understand the acoustic field effect on the coupling of droplet characteristics (such as droplet size, droplet velocity, and number density of droplet) and combustion instability. Recently, Arad et al. \cite{arad2023} numerically (using ANSYS Fluent) investigated the effect of standing acoustic waves on a stream of droplets and found that acoustic waves significantly affect droplet grouping and droplet distribution. 

It is evident from the present literature review that not enough studies have been done to understand the effects of acoustic waves on droplet grouping and coalescence. The present work is a first step in an ongoing effort to fill this research gap. This work presents an experimental investigation in order to understand the influence of a standing acoustic wave on a single stream of droplets.

\begin{figure}
\centering
\includegraphics[width=0.8\linewidth]{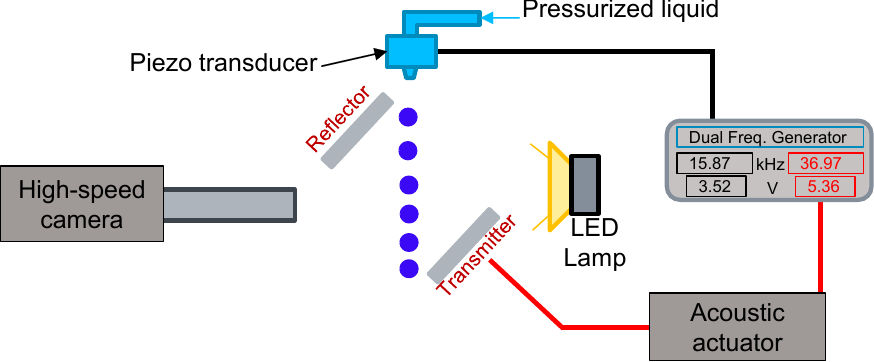}
\caption{Experimental setup}
\label{setup}
\end{figure}

\begin{figure}
\centering
\includegraphics[width=0.5\linewidth]{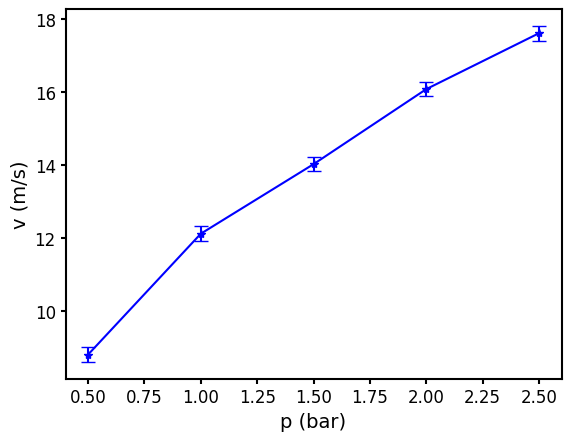}
\caption{Droplet stream velocity versus liquid pressure of nozzle.}
\label{flow}
\end{figure}

\section{Experimental setup}

Figure \ref{setup} shows the experimental setup used in the present study. The experimental setup has three major parts: a high-speed visualization setup, the droplet stream generator, and the acoustic field apparatus. The high-speed visualization of the stream of droplets was done using a Photron SA1.1 high-speed camera, and the background was illuminated using a LED lamp. The droplet grouping phenomena were captured at 10,000 frames per second. The acoustic field generator configuration comprises an acoustic transmitter (of a resonant frequency of 36.97 kHz) and a reflector plate, allowing the droplet stream to pass through in the presence or absence of an external pressure field generated by a standing acoustic wave. The transmitter is excited through an acoustic actuator, which receives signal from a dual-frequency generator. Using the frequency generator, amplitude, waveform, and frequency of the supplied signal can be controlled. In the present work, a sinusoidal waveform with a 5.36 V amplitude and 36.97 kHz frequency was supplied to the transmitter using the acoustic actuator. To generate a stream of droplets, an orifice/nozzle of 125 $\mu$m diameter ($d_n$) was used, which was attached to a piezo transducer. As shown in Figure \ref{setup}, the pressurized liquid (Isopropanol) is supplied to the nozzle in the range of 0.5 bar to 2.5 bar. By controlling the liquid pressure, the flow rate through the nozzle was controlled. Figure \ref{flow} shows the mean flow velocity with respect to the injection pressure of the liquid ($p$). The mass flow rate or flow velocity increased as the injection pressure of the liquid ($p$) increased. Corresponding Reynolds numbers (Re = $\rho_l v d_n/\mu_l$, subscript $l$ represents liquid properties) for pressure values $p$ = 0.5, 1.0, 1.5, 2.0, and 2.5 bar are 702, 970, 1122, 1286, and 1409, respectively. As the liquid stream emerges from the nozzle, it breaks into small droplets due to vibrations produced using the piezo transducer based on the input excitation frequency ($f$) supplied to it. The excitation signal is provided by the other connection of the dual frequency generator, and as explained earlier, the amplitude, waveform, and frequency of this signal can also be controlled. In the present work, the amplitude was kept constant at 3.52 Volts, while the excitation frequency was varied from 3.4 kHz to 50 kHz.

\section{Results}

\begin{figure}
\centering
\includegraphics[width=0.7\linewidth]{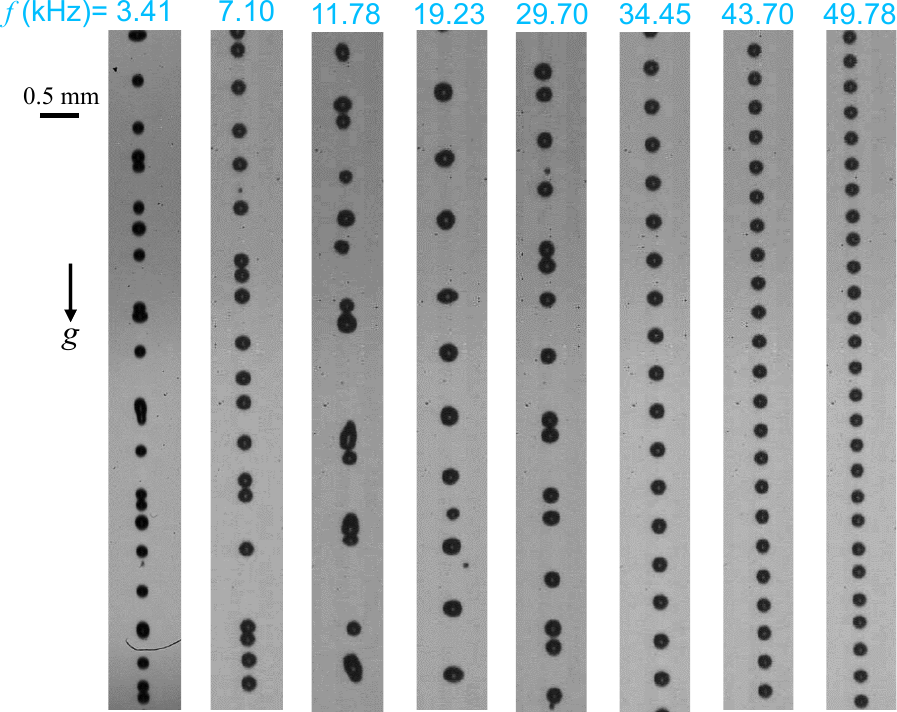}
\caption{Effect of different piezo frequencies on the droplet stream at a fixed 
injection pressure of 1.5 bar and Re = 1122.}
\label{freqeffec}
\end{figure}

\subsection{Effect of piezo frequency on droplet stream}
Figure \ref{freqeffec} shows the effect of the piezo frequency on the droplet stream generated by a nozzle of 125 $\mu$m diameter at a constant pressure ($p$) of 1.5 bar and Re = 1122, without the acoustic field. The figure shows the snapshots of the droplet stream recorded by a high-speed camera at eight different excitation frequencies ($f$) when a steady stream was observed in each case. At lower frequencies (i.e., 3.41, 7.10, 11.78, 19.23, 29.70 kHz), the droplet stream shows an irregular behavior. Within the stream, the spacing between the droplets is not uniform, and also the size of the droplets is not the same. However, it can also be observed that, as the frequency increases from 3.41 kHz to 29.70 kHz, the droplet stream starts to organize itself in terms of size and spacing. After a critical piezo-frequency (between 29.70 and 34.45 kHz), the droplet stream gets organized fully and shows the uniformity of size and spacing, as it can be seen in Figure \ref{freqeffec} for the frequencies 34.45 to 49.78 kHz. Also, the spacing between the droplets and droplet size decreases as the frequency increases from 34.45 to 49.78 kHz.

\begin{figure}
\centering
\includegraphics[width=0.55\linewidth]{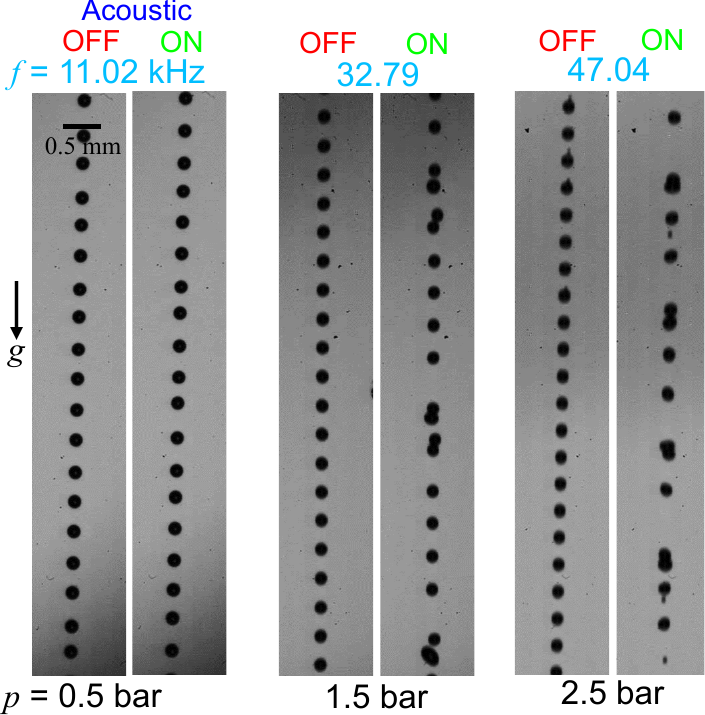}
\caption{Effect of acoustics on the droplet stream at different injection pressures 
and piezo-frequencies. The Reynolds numbers (Re) for pressure $p$ = 0.5, 1.5 and 2.5 bar are 702, 1122, and 1409, respectively.}
\label{pnfeffec}
\end{figure}

\subsection{Effect of acoustics on droplet stream}
Figure \ref{pnfeffec} shows the effect of acoustics on the droplet stream at different injection pressures and piezo-frequencies in the absence and presence of the acoustic field. The first observation can be made from this figure, even at the low frequency of 11.02 kHz, the droplet stream is uniform. This can be attributed to the low pressure of 0.5 bar, at which low-frequency vibrations are sufficient to break up the jet stream coming out of the nozzle. At a low pressure of 0.5 bar and a low frequency of 11.02 kHz, no effect of the acoustic field was observed. However, at higher pressures (i.e., 1.5 and 2.5 bar) and higher frequencies (i.e., 32.79 and 47.04 kHz), on turning ON the acoustic field, droplets in the droplet stream start to group and coalesce with each other, and in this process losing its uniform nature. The presence of an acoustic field creates a pressure field, which helps in bringing droplets closer to each other. Also, increasing the frequency results in a higher number of grouping events and eventual coalescence in the presence of the acoustic field. This happens due to less spacing between the droplets at higher frequencies.

\begin{figure}
\centering
\includegraphics[width=0.75\linewidth]{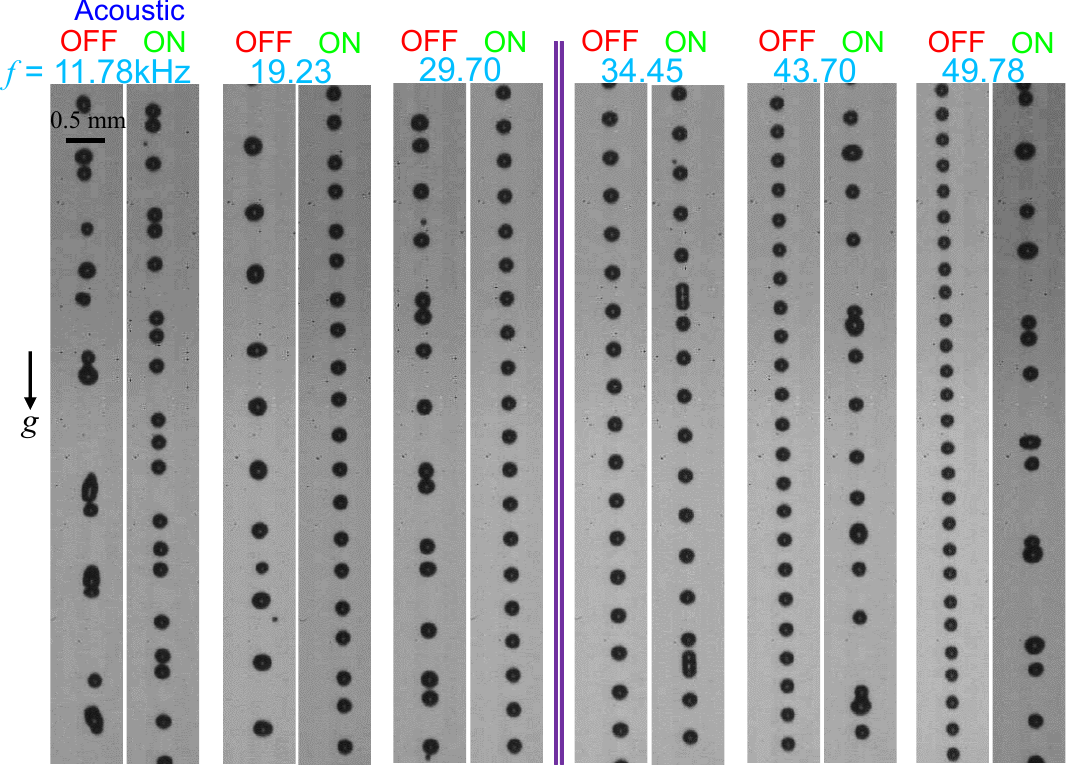}
\caption{Competing effects of piezo-frequency and acoustic field on droplet stream. $p$ = 1.5 bar, Re = 1122.}
\label{compeffec}
\end{figure}

\subsection{Competing effects of frequency and acoustic field}
The competing effect of nozzle excitation frequency and acoustic field can be observed in Figure \ref{compeffec}. At lower piezo frequencies (i.e., 11.78, 19.23, and 29.70 kHz), an unstable stream of droplets with different sizes and spacings between them is generated by the nozzle. When the acoustic field is applied for these lower frequencies, the stream of droplets becomes organized, and in some cases, it becomes equispaced and of the same size. Also, it can be observed that larger droplets break into smaller droplets when the acoustic field is turned ON. However, an opposite behavior has been observed at higher frequencies (i.e., 34.45, 43.70, 49.78 kHz). In these cases, an equispaced stream of same-size droplets group and coalesce with each other and becomes non-uniform as the acoustic field is applied.

\section{Conclusions}

The present work experimentally investigates the effect of an acoustic field on a stream of droplets. The droplet stream was generated at different injection pressures ranging from 0.5 to 2.5 bar with different droplet generator piezo-frequencies (from 3.41 to 50 kHz). The droplet size and spacing between them were controlled by controlling pressure and frequency. An acoustic transmitter of the resonant frequency of 36.97 kHz was employed to observe the effect of an acoustic field. It was observed from this study that the acoustic field can enhance the grouping and coalescence of droplets in a stream of droplets in the case of higher piezo-frequencies. However, at lower piezo-frequencies, the application of an acoustic field may result in the breakup of droplets into smaller droplets.

\section*{Acknowledgement}
This study was conducted with the financial support of the Deutsche Forschungsgemeinschaft (DFG, German Research Foundation) through the project 409029509, "Investigation of Droplet Motion and Grouping".

\section*{References}

\bibliography{ManishBib}

\end{document}